# FinRetrieval: A Benchmark for Financial Data Retrieval by AI Agents

Eric Y. Kim[1]   Jie Huang[1]

[1]Daloopa

January 2026

**Abstract**

AI agents increasingly assist with financial research, yet no benchmark evaluates their ability to retrieve specific numeric values from structured databases. We introduce FinRetrieval, a benchmark of 500 financial retrieval questions with ground truth answers, agent responses from 14 configurations across three frontier providers (Anthropic, OpenAI, Google), and complete tool call execution traces. Our evaluation reveals that tool availability dominates performance: Claude Opus achieves 90.8% accuracy with structured data APIs but only 19.8% with web search alone—a 71 percentage point gap that exceeds other providers by 3–4×. We find that reasoning mode benefits vary inversely with base capability (+9.0pp for OpenAI vs +2.8pp for Claude), explained by differences in base-mode tool utilization rather than reasoning ability. Geographic performance gaps (5.6pp US advantage) stem from fiscal year naming conventions, not model limitations. We release the dataset, evaluation code, and tool traces to enable research on financial AI systems.

## 1 Introduction

AI agents are increasingly deployed for financial research tasks: answering questions about company performance, retrieving specific metrics from earnings reports, and synthesizing data across sources. These applications require agents to accurately retrieve numeric values from structured databases—a capability that existing benchmarks do not directly evaluate.

Current financial QA benchmarks focus on numerical reasoning over provided documents. FinQA [1] tests arithmetic calculations given table excerpts. TAT-QA [2] evaluates hybrid reasoning over tabular and textual data. FinanceBench [3] provides open-book questions but does not evaluate agent tool use. Recent benchmarks such as FAB [4] have begun evaluating tool-augmented agents across multiple providers, but focus on document retrieval from SEC filings. No existing benchmark evaluates retrieval from structured financial databases.

We introduce **FinRetrieval**, a benchmark specifically designed to evaluate AI agents on financial data retrieval. The benchmark comprises 500 questions targeting specific numeric values (e.g., "What was Apple's Q3 2024 revenue?"), ground truth answers verified against source documents, and agent responses from 14 configurations spanning three frontier providers (Anthropic, OpenAI, Google). Critically, we release complete execution traces (every tool call with inputs and outputs), enabling detailed analysis of agent decision-making.

Our evaluation reveals four findings with practical implications:

1. **Tool availability has over 3× larger impact than model selection.** Claude Opus drops from 90.8% to 19.8% accuracy without structured data tools, while Google drops only from 90.6% to 69.2%. The 71pp gap for Claude suggests that tool integration, not model capability, is the primary determinant of performance.

2. **Reasoning mode benefits vary inversely with base capability.** OpenAI gains +9.0pp from extended reasoning while Claude Opus gains only +2.8pp. Analysis of tool call traces reveals that OpenAI's base mode makes fewer API exploration calls, and reasoning compensates for this gap.

3. **First-query success drives efficiency, not vice versa.** Correct answers correlate with fewer tool calls, but this reflects question difficulty—first-query hits yield 93% accuracy versus 77% for questions requiring additional searches.

4. **Geographic gaps stem from data conventions.** The 5.6pp US vs non-US accuracy gap for Claude is driven by fiscal period mismatches, not model limitations. At the December fiscal year-end baseline, non-US accuracy exceeds US accuracy.

We release the complete dataset including questions, responses, tool traces, and evaluation code to enable reproducible research on financial AI systems.

The remainder of this paper is organized as follows. Section 2 surveys related work in financial QA and agent evaluation. Section 3 describes the benchmark design and question generation pipeline. Section 4 details the experimental setup including models, tools, and scoring methodology. Section 5 presents our main results. Section 6 provides error analysis. Section 7 discusses implications and limitations, and Section 8 concludes.

## 2 Related Work

Our work builds on two research streams: financial question answering benchmarks and evaluations of tool-augmented language models.

### 2.1 Financial Question Answering

**Document-based reasoning.** FinQA [1] pioneered numerical reasoning over financial tables, requiring models to perform arithmetic calculations given table excerpts from earnings reports. TAT-QA [2] extended this to hybrid tabular and textual contexts, testing reasoning across both modalities. DocFinQA [5] scaled to longer documents, evaluating comprehension over full SEC filings. These benchmarks test **reasoning** over provided information but do not evaluate the ability to **retrieve** that information in the first place.

**Open-book financial QA.** FinanceBench [3] introduced open-ended questions about public company financials, allowing models to access external sources. FinBen [6] provides holistic evaluation across diverse financial NLP tasks. FinTextQA [7] provides long-form financial QA pairs curated from finance textbooks. While these benchmarks move toward realistic retrieval scenarios, they do not evaluate agent tool use or compare multiple providers on identical questions.

**Retrieval-augmented financial QA.** ChatLR [8] approaches structured database retrieval by fine-tuning LLMs for Text2API generation, achieving high accuracy but requiring model fine-tuning rather than evaluating general-purpose agents. FinDER [9], FinAgentBench [10], and Finance Agent Benchmark [4] evaluate retrieval over financial documents, with FinAgentBench introducing multi-step agentic reasoning and FAB providing tool access for SEC filing analysis. FAMMA [11] adds multilingual and multimodal dimensions. None of these benchmarks combine structured API access with general-purpose agents, and none release complete execution traces for behavioral analysis.

## 2.2 Tool-Augmented Language Models

**General tool use benchmarks.** API-Bank [12] provides a comprehensive benchmark for tool-augmented LLMs across diverse API categories. StableToolBench [13] addresses stability and reproducibility concerns in tool learning evaluation. AgentBench [14] evaluates LLMs as agents across eight diverse environments including operating systems, databases, knowledge graphs, and web tasks. These benchmarks establish methodologies for tool use evaluation but operate in general domains rather than specialized financial contexts.

**Retrieval-augmented generation.** CRAG [15] provides comprehensive evaluation of RAG systems across diverse question types. The ReAct framework [16] demonstrates how reasoning and acting can be synergized in language models. Survey work on augmented language models [17] provides a comprehensive taxonomy of tool integration approaches. However, these works focus on web retrieval rather than structured financial databases.

**Reasoning in LLMs.** Chain-of-thought prompting [18] and extended reasoning modes like DeepSeek-R1 [19] have shown substantial improvements in complex reasoning tasks. Our Finding B contributes to this literature by showing that reasoning benefits vary with base-mode tool utilization quality.

## 2.3 Our Contribution

FinRetrieval differs from prior work in three ways. First, we evaluate **structured API access** via the Model Context Protocol rather than document retrieval or web search. Second, we compare **multiple frontier providers** (Anthropic, OpenAI, Google) on identical questions with controlled tool configurations. Third, we release **complete execution traces**—every tool call with inputs and outputs—enabling detailed analysis of agent decision-making that prior benchmarks do not support.

# 3 Benchmark Design

FinRetrieval is designed to evaluate AI agents on financial data retrieval—the task of finding and extracting specific numeric values from structured databases given natural language questions. This section describes our question generation pipeline, dataset composition, and data quality controls.

Questions and ground truth answers are drawn from Daloopa, a financial data platform that extracts structured data from SEC filings, earnings releases, and investor presentations for over 4,000 public companies. Data points are analyst-curated and linked to source documents, providing traceable ground truth for evaluation. The authors are employed by Daloopa; limitations of this approach are discussed in Section 7.3.

## 3.1 Question Generation Pipeline

Questions are generated through a three-phase pipeline that ensures each question has an unambiguous ground truth answer tied to a verifiable source document.

**Phase 1: Stratified Sampling.** We sample data points through a four-stage process: (1) selecting a company with 50:50 US/non-US stratification, (2) selecting one of six financial categories uniformly at random, (3) selecting a metric series using hierarchical weighted sampling within the chosen company-category subset, and (4) uniformly sampling from available data points for that series. Categories are assigned by pattern-matching the top-level hierarchy of each series name (e.g., "Income Statement | Revenue | Total" maps to income statement performance). The series sampling uses $\sqrt{\log(n)}$ weighting to reduce over-sampling of "breakdown" metrics (see Appendix E for details).

**Phase 2: Context Enrichment.** For each sampled data point, we retrieve two forms of context: (1) adjacent series ($\pm 10$ rows capturing hierarchical context: totals, subtotals, and related line items), and (2) a screenshot of the source document with the relevant cell automatically highlighted. The screenshot serves as the authoritative source for period disambiguation, showing explicitly whether values represent "three months ended" versus "year ended."

**Phase 3: Question Generation.** A frontier LLM (GPT-5.1) generates question-answer pairs from a multimodal prompt combining visual context (screenshot as primary source), structured data reference (company, metric, period, value), and task instructions (disambiguation rules, sign conventions, period naming requirements). This visual-centric design grounds questions in source document evidence rather than database field labels alone. The model produces structured output containing a natural language question and cited answer.

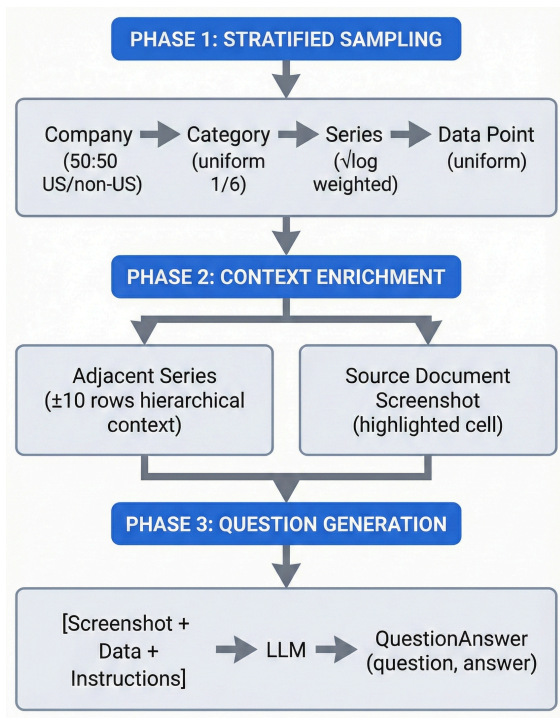

Figure 1: Question generation pipeline. Phase 1 samples data points through stratified selection. Phase 2 enriches each sample with hierarchical context and source document evidence. Phase 3 generates question-answer pairs via multimodal LLM.

| Category | Questions | % |
| --- | --- | --- |
| Balance Sheet | 98 | 19.6% |
| Cash Flow | 93 | 18.6% |
| Operational KPIs | 92 | 18.4% |
| Income Statement | 80 | 16.0% |
| Guidance/Outlook | 73 | 14.6% |
| Segments/Geography | 64 | 12.8% |
| **Total** | **500** | **100%** |

Table 1: Dataset distribution by financial category.

## 3.2 Dataset Composition

Questions are drawn from a sampling pool of 4,649 public companies. Data points span fiscal years 2015 onwards.

The final dataset comprises 500 questions spanning six financial categories (Table 1). Representative examples from each category are provided in Appendix A.

Period distribution is 59% quarterly, 33% annual, and 8% half-year. Geographic distribution is 271 US questions (54.2%) and 229 non-US (45.8%)[1], drawn from companies across 41 countries. The non-US subset includes Japan (10%), Australia, the United Kingdom, and India.

## 3.3 Data Quality Filters

To ensure every question has a verifiable ground truth, we exclude two classes of data:

**Auto-generated entries.** Values tagged automatically without human verification, which have higher error rates and inconsistent units.

**Computed values.** Figures derived through calculation rather than sourced directly from filings—for example, second-half results inferred from annual minus first-half, or quarterly breakdowns for companies reporting only semi-annually. These entries lack source document screenshots.

These filters remove approximately 9% of the database while ensuring every answer traces to a specific cell in a source document.

# 4 Experimental Setup

We evaluate three frontier LLM providers across multiple configurations, varying tool availability and reasoning mode. This section describes our model selection, tool configuration, and scoring methodology.

## 4.1 Models and Configurations

We evaluate agents from three providers: Anthropic, OpenAI, and Google. Each provider offers both a base configuration and an extended reasoning mode. Additionally, we test with and without access to structured financial data tools, yielding 14 configurations total (Table 2).

---

[1] US includes companies traded on US exchanges, including ADRs of non-US headquartered companies.

| Model | Web | Web+R | MCP | MCP+R |
|---|---|---|---|---|
| Claude Opus 4.5 | ✓ | ✓ | ✓ | ✓ |
| Claude Sonnet 4.5 | — | — | ✓ | ✓ |
| GPT-5.2 | ✓ | ✓ | ✓ | ✓ |
| Gemini 3 Pro | ✓ | ✓ | ✓ | ✓ |

Table 2: The 14 evaluation configurations by model and tool/reasoning mode. Web = WebSearch only; MCP = structured financial data API; +R = reasoning enabled. Sonnet was not evaluated in WebSearch-only mode.

The Sonnet model (Anthropic's smaller variant) is included only in MCP configurations to enable within-provider tier comparison. All agents are implemented using their respective official SDKs: Claude Agent SDK, OpenAI Agents SDK, and Google ADK. Agent responses and tool call traces are normalized to a common output format regardless of SDK-specific interfaces.

**Agent-as-a-whole evaluation.** This benchmark evaluates complete agent systems rather than isolated model capabilities. Each configuration comprises the provider's full stack: model, agent SDK, and native web tools. This establishes a baseline for vendor-provided agent stacks as practitioners would deploy them; custom harnesses (LangChain, LlamaIndex, etc.) may yield different results. We control for task (500 identical questions), structured data source (identical MCP tools), and evaluation criteria. Cross-provider performance differences reflect complete systems; within-provider comparisons (e.g., MCP vs WebOnly, reasoning vs base) isolate specific factors and support stronger causal claims.

## 4.2 Tool Configuration

Agents operate in one of two tool modes:

**MCP Mode.** Agents have access to web search plus three structured data tools via the Model Context Protocol (MCP):

- `discover_companies`: Maps company names or ticker symbols to internal identifiers
- `discover_company_series`: Returns available financial metrics for a company, filtered by keywords
- `get_company_fundamentals`: Retrieves specific values for given company, metrics, and time periods

These tools provide direct access to a structured financial database containing data from SEC filings, earnings releases, and investor presentations for over 4,000 public companies. Complete MCP tool schemas and example execution traces are included in the released code and dataset.

**WebOnly Mode.** Agents have access only to web search, simulating a scenario where structured financial APIs are unavailable. This baseline tests agents' ability to find financial data through general web retrieval.

All configurations include web search capability. The web search implementation varies by provider: Google Search for Google ADK, and provider-specific web search tools for Claude and OpenAI.

## 4.3 Reasoning Mode

Each provider offers an extended reasoning capability, though implementations differ:

- **Claude**: Extended thinking with `max_thinking_tokens=8000`
- **OpenAI**: Reasoning API with `reasoning_effort="high"`

- **Google**: ThinkingConfig with `thinking_level="high"` (reasoning cannot be fully disabled; base mode uses "low")

These differences mean cross-provider reasoning comparisons should be interpreted cautiously—the +9.0pp benefit for OpenAI versus +2.8pp for Claude Opus (Finding B) may partially reflect implementation differences rather than pure model capability. Detailed parameters are provided in Appendix F.

### 4.4 Scoring

Responses are scored by an LLM judge (GPT-5.2 with temperature 0) using binary correctness. The judge receives the question, expected answer, and agent response, then determines whether the response contains a value matching the expected answer.

The scoring prompt includes normalization rules:
- **Unit equivalence**: "$1.5 billion" matches "1,500 million" matches "1500000 thousand"
- **Sign tolerance**: Parentheses indicate negative values; context words like "loss" or "decrease" may indicate sign

A response is marked correct if the agent provides the expected numeric value with appropriate context. Responses that decline to answer, provide wrong values, or cite incorrect time periods are marked incorrect. The full scoring prompt is provided in Appendix C.

**Scorer validation.** To identify false negatives (correct responses incorrectly marked wrong), we manually reviewed all 524 responses initially marked incorrect across the eight MCP-enabled configurations. Each failure was analyzed using a structured root cause workflow examining tool call traces, agent reasoning, and MCP outputs. This identified 13 scorer errors (2.5% of failures), primarily involving sign representation ambiguity, rounding tolerance inconsistency, and value extraction from multi-format responses. All were corrected via manual override. False positive rate (incorrect responses marked correct) was not estimated; systematic sampling of "correct" responses remains future work.

## 5 Results

We evaluate 14 agent configurations across 500 financial retrieval questions, producing 7,000 total evaluations. This section presents our main findings on accuracy, computational cost, tool availability impact, reasoning mode benefits, tool call efficiency, and geographic performance gaps.

### 5.1 Overall Accuracy

Table 3 summarizes accuracy across all 14 configurations. The best-performing configuration (Claude Opus with reasoning and MCP tools) achieves 90.8% accuracy, while the worst (Claude Opus with reasoning but WebSearch only) achieves just 19.8%—a 71 percentage point spread.

Three patterns emerge from these results. First, all MCP-enabled configurations achieve 80–91% accuracy, while WebSearch-only configurations range from 19.8–70.8%. Second, reasoning mode provides modest gains within the MCP setting (+2.8 to +9.0pp) but does not compensate for lack of structured data access. Third, Claude shows the largest sensitivity to tool availability, with accuracy swinging 71 percentage points between its best and worst configurations.

| Model | Web | Web+R | MCP | MCP+R |
|---|---|---|---|---|
| Claude Opus 4.5 | 23.0% | **19.8%** | 88.0% | **90.8%** |
| Claude Sonnet 4.5 | — | — | 81.8% | 86.0% |
| GPT-5.2 | 51.2% | 70.8% | 80.2% | 89.2% |
| Gemini 3 Pro | 64.0% | 69.2% | 86.6% | 90.6% |

Table 3: Accuracy by model and configuration. Web = WebSearch only; MCP = structured financial data tools; +R = reasoning mode enabled. Bold indicates best and worst performers. Sonnet was not evaluated in WebSearch-only mode. 95% CIs range ±2.6pp (at 90%) to ±4.0pp (at 70%); pairwise differences >5pp are significant at $p < 0.05$ (Appendix H).

| Model | Base | +Reasoning | Overhead |
|---|---|---|---|
| GPT-5.2 | 9.9s | 31.2s | **3.2×** |
| Claude Sonnet 4.5 | 19.6s | 25.0s | 1.3× |
| Claude Opus 4.5 | 21.2s | 25.5s | 1.2× |
| Gemini 3 Pro | 36.9s | 44.5s | 1.2× |

Table 4: Median response latency by model and reasoning mode (MCP-enabled configurations). Ordered by reasoning overhead.

## 5.2 Computational Cost

Beyond accuracy, deployment decisions depend on response time and resource usage. Table 4 reports median latency across MCP-enabled configurations.[2]

Reasoning mode latency overhead varies substantially across models. OpenAI has the lowest base latency (9.9s) but reasoning mode increases it 3.2× to 31.2s—higher than any other configuration. Claude and Google show consistent ~1.2× overhead regardless of base latency. The result is a convergence: with reasoning enabled, all models require 25–45 seconds per query despite 4× differences in base latency.

Token consumption increases modestly with reasoning (1.0–1.2×) across all models. Claude configurations consume substantially more tokens (~75,000 median) than OpenAI (~28,000) or Google (~9,000), reflecting differences in how extended thinking tokens are reported. Token overhead is not the primary cost driver; latency variance dominates.

The interaction between accuracy gains (Section 5.4) and latency costs is analyzed in Section 7.1.

## 5.3 Finding A: Tool Availability Impact

The most dramatic finding concerns tool availability. When equipped with MCP tools providing structured access to financial data, Claude Opus achieves 90.8% accuracy (95% CI: 88.3–93.3%). Without these tools—relying on WebSearch alone—accuracy drops to 19.8% (95% CI: 16.3–23.3%), a **71 percentage point gap** that is highly significant ($z = 22.5$, $p < 0.001$). This exceeds the tool availability impact for other models by a factor of 3–4× (Figure 2).

---

[2]Latencies measured from a residential fiber connection; at 10–45 second response times, local hardware and network contribute negligible variance compared to provider-side processing.

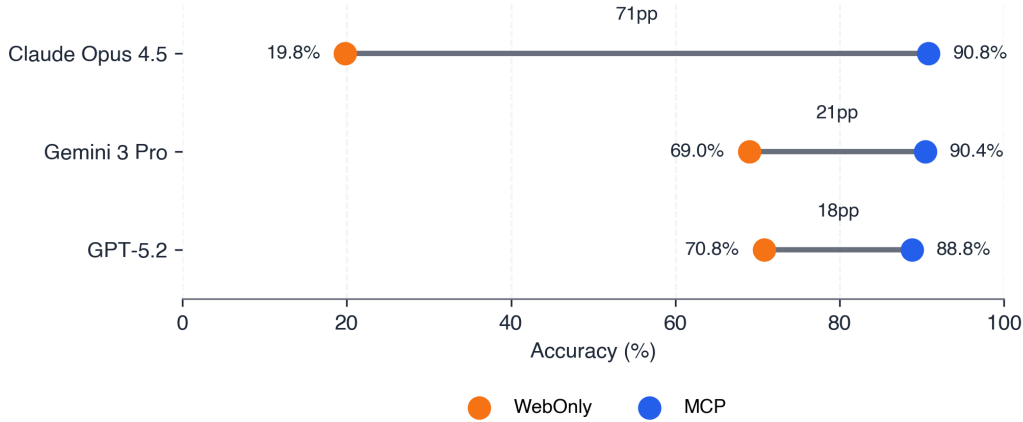

Figure 2: Tool availability impact by model (reasoning mode enabled). Claude's 71pp gap is 3–4× larger than GPT or Gemini.

The gap decomposition reveals two components. First, a **behavioral component**: Claude's WebSearch-only configuration exhibits a 55% "gave up" rate—cases where the agent found relevant information but failed to commit to an answer, instead continuing to search until timing out or explicitly declining to answer. By comparison, Google's gave-up rate is under 1%. This behavioral pattern accounts for approximately 35pp of Claude's gap.

Second, an **infrastructure component**: Claude's WebSearch tool provides only search result snippets, not full document access or PDF parsing. Google's and OpenAI's web tools can retrieve full pages and parse PDF tables from SEC filings. This infrastructure gap amplifies Claude's behavioral tendency toward extensive searching.

We provide detailed case studies of the behavioral pattern in Appendix G, including five representative questions where Claude found the correct answer in search results but failed to extract and commit to it.

### 5.4 Finding B: Reasoning Benefits Vary by Model

Extended reasoning provides varying benefits across models (Table 5).

The root cause is not that reasoning helps OpenAI more, but that OpenAI's base mode has weaker MCP tool utilization. Analysis of tool call traces reveals that OpenAI's base mode makes an average of 1.03 `discover_company_series` (DS) calls per question, compared to 1.78 for Claude Opus—73% more exploration. OpenAI's reasoning mode compensates for this gap through more careful series selection from MCP output, not through additional exploration (DS calls increase only to 1.18, a 15% gain).

| Model | Base | Reasoning | Delta |
|---|---|---|---|
| GPT-5.2 | 80.2% | 89.2% | **+9.0pp** |
| Claude Sonnet 4.5 | 81.8% | 86.0% | +4.2pp |
| Gemini 3 Pro | 86.6% | 90.6% | +4.0pp |
| Claude Opus 4.5 | 88.0% | 90.8% | +2.8pp |

Table 5: Reasoning mode benefit by model (MCP-enabled configurations). OpenAI shows 3× larger benefit than Claude Opus.

| Model | DS Success Rate | Acc if Success | Acc if Failure |
|---|---|---|---|
| Claude Opus 4.5 | 87.2% | 92.9% | 76.6% |
| Gemini 3 Pro | 81.0% | 91.6% | 85.3% |
| GPT-5.2 | 73.2% | 92.3% | 79.1% |

Table 6: First DS call success rate and resulting accuracy. The 6–16pp accuracy gap between success and failure explains the tool count correlation.

This finding suggests that reasoning mode benefits are largest for models with weaker base-mode tool utilization. All four models converge to 86–91% accuracy with reasoning enabled, indicating a ceiling determined by benchmark difficulty and tool design rather than model capability.

### 5.5 Finding C: Tool Call Efficiency

Correct answers correlate with fewer tool calls: Claude Opus averages 3.5 MCP calls for correct answers versus 5.5 for incorrect ones (37% fewer). However, this correlation is **not causal**—it is mediated by success or failure on the first `discover_company_series` (DS) call.

The optimal retrieval path requires exactly three MCP calls:
1. `discover_companies` — map ticker to company ID
2. `discover_company_series` — find relevant financial metrics
3. `get_company_fundamentals` — retrieve values for specific periods

When the first DS call returns the target series (87% of cases for Claude), agents complete this 3-call path with 92.9% accuracy. When it fails, agents enter extended search loops, making 4–18 additional calls with accuracy dropping to 76.6%—a **16.3pp gap** (Table 6).

What predicts first DS success? Financial category has a 16pp range: cash flow questions achieve 95% first-DS success versus 79% for operational KPIs. Series hierarchy depth, initially hypothesized as a predictor, has less than 1pp effect. Re-querying 58 first-DS failures without period constraints reveals two distinct bottlenecks: 55% (32/58) are keyword matching failures where the target series is not found even without period filtering, while 45% (26/58) are period filter failures where the target is found when the period constraint is removed.

Root cause analysis of the 26 period filter failures identifies three patterns: **period convention misalignment** (46%), encompassing both quarter identity differences (fiscal Q1 $\neq$ calendar Q1 for non-December FYE companies) and year labeling conventions (start-year vs end-year)—the same underlying issue that drives the geographic accuracy gap (Section 5.6); **granularity mismatch** (27%), where agents query annual periods for data stored at quarterly granularity; and **guidance storage conventions** (23%), where guidance is indexed by announcement period rather than target period. Period convention misalignment is addressable via API changes (supporting both formats) or documentation with fiscal year-end data for agent-side conversion. Granularity and guidance issues are addressable via documentation.

### 5.6 Finding D: Geographic Performance Gap

All models show higher accuracy on US companies than non-US companies (Table 7).

| Model | US (n=271) | Non-US (n=229) | Gap |
|---|---|---|---|
| Claude Opus 4.5 | 93.4% | 87.8% | +5.6pp |
| Gemini 3 Pro | 92.6% | 88.2% | +4.4pp |
| GPT-5.2 | 90.4% | 87.8% | +2.6pp |

Table 7: US vs non-US accuracy (reasoning mode, MCP enabled). The gap is driven by fiscal year naming conventions, not geography.

The gap is **not** inherent to geography. At the December fiscal year-end baseline (standard calendar alignment), non-US accuracy (95.8%) slightly exceeds US accuracy (95.6%). The gap emerges from companies with non-standard fiscal year-ends:

- **March FYE** (common for Japanese companies): 64.7% accuracy (17 questions)
- **September FYE** (common for Indian companies): 78.6% accuracy (42 questions)
- **Unknown FYE**: 77.8% for non-US (27 questions) vs 92.9% for US (14 questions)

The root cause is period convention misalignment (the same issue identified in period filter failures) (Section 5.5). For annual queries, this manifests as year labeling mismatch: our data source uses fiscal year **start** year (e.g., "2022FY" covers April 2022–March 2023), while agents assume **end** year (querying "2023FY"). For quarterly queries, fiscal quarters map to different calendar months entirely. Both patterns disproportionately affect non-December FYE companies.

Non-US companies have 2× higher fraction of questions in problematic segments (48% vs 24% in non-standard FYE or unknown fiscal year-end). The gap is thus an artifact of data conventions, not model capability or training data distribution.

# 6 Error Analysis

To understand why agents fail, we analyzed errors at two levels. First, we categorized all 524 incorrect responses across eight MCP-enabled configurations by high-level error type (Appendix I). Second, for the 46 agent errors from the best-performing configuration (Claude Opus with reasoning), we conducted detailed root cause analysis tracing each error to specific tool calls and reasoning steps.

## 6.1 Root Cause Distribution

The detailed analysis of Claude Opus with reasoning failures reveals a concentrated distribution (Table 8).

| Root Cause | Count | % |
|---|---|---|
| Period confusion | 29 | 63.0% |
| Wrong series selection | 9 | 19.6% |
| Other agent errors | 7 | 15.2% |
| Sign errors | 1 | 2.2% |
| **Total** | **46** | **100%** |

Table 8: Root cause distribution for Claude Opus with reasoning failures (best-performing MCP configuration at 90.8% accuracy).

Period confusion alone accounts for nearly two-thirds of all failures. This category encompasses period convention misalignment (both quarter identity differences and year labeling conventions), granularity mismatches (FY vs Q4), and guidance storage periods (announcement date vs target date). The pattern is systematic: agents understood the questions correctly but queried incorrect periods due to undocumented tool conventions.

Wrong series selection (20%) occurs when agents retrieve similar but incorrect metrics, such as "gross production" instead of "net production," or selecting a sub-component when the total was requested.

Other agent errors (15%) include field selection (quarterly vs YTD values), search strategy failures, and data interpretation issues.

### 6.2 Representative Case Study

To illustrate period confusion, consider a question about fiscal year ended March 2023 revenue. The agent queried `2023FY`, but Daloopa uses starting-year convention where April 2022–March 2023 is labeled `2022FY`. The agent retrieved data from the wrong year entirely.

This isn't a reasoning failure—the agent understood the question correctly. It failed because the tool's period naming convention wasn't documented clearly enough for the model to infer the correct mapping. Better tool documentation could address nearly two-thirds of current errors.

A high-level error taxonomy across all configurations is provided in Appendix I. Appendix A provides additional period confusion examples illustrating common agent mistakes.

# 7 Discussion

We discuss implications for practitioners, identify paths to higher accuracy, and acknowledge limitations.

### 7.1 Implications for Practitioners

**Tool capabilities matter more than model selection.** With structured data APIs, provider choice barely matters—all achieve 89–91% accuracy. Without structured data, web search tool quality becomes critical: Claude's snippet-only access yields 20% while OpenAI and Google achieve ~70% with full-page browsing. Practitioners should prioritize structured data integration; when unavailable, evaluate web search capabilities carefully.

**Reasoning mode ROI varies.** OpenAI gains +9.0pp accuracy from reasoning but pays 3.2× latency overhead; Claude Opus gains +2.8pp at only 1.2× overhead (Section 5.2). Per unit latency increase, OpenAI gains 3× less accuracy than Claude or Google. For models with strong base-mode tool utilization, the marginal accuracy gain may not justify the latency cost. Practitioners should benchmark both accuracy and latency for their specific use case.

**Data conventions require explicit handling.** The 5.6pp US vs non-US accuracy gap, driven by fiscal year naming conventions, demonstrates that financial AI systems must explicitly handle regional data formatting differences.

## 7.2 Path to Higher Accuracy

Our analysis identifies the primary levers for improving financial retrieval accuracy:

**Tool availability is the dominant factor.** The tool availability gap (Section 5.3) dwarfs the impact of reasoning mode (+2.8 to +9.0pp) or model selection (1.6pp spread among MCP+reasoning configurations). Expanding tool coverage should be the first priority.

**Undocumented tool conventions block the path beyond 90%.** With structured tools, accuracy reaches 87–91%. For the best-performing configuration, remaining errors concentrate on period confusion (63% of failures), caused by undocumented fiscal year labeling conventions—start-year vs end-year naming. Adding explicit convention details to tool descriptions could address the majority of current failures without model changes.

**First-query success is critical.** Finding C reveals that 93% accuracy is achievable when the first `discover_company_series` call succeeds, versus 77% when it fails. Two interventions could reduce first-query failures: (1) API-side fuzzy matching and synonym expansion for keyword mismatches ( 55% of failures), and (2) period convention alignment ( 45% of failures)—46% stem from period convention misalignment (addressable via API support for both formats or fiscal year-end data enabling agent conversion), while 50% are addressable via documentation (see Section 5.5).

## 7.3 Limitations

**Single data source.** FinRetrieval uses a single financial database (Daloopa) as both the data source for question generation and the target for agent retrieval. Performance may not generalize to other financial data providers with different schemas, coverage, or API designs.

**Single-number retrieval only.** All questions request a single data point with a known answer. Tasks requiring multi-step calculations, synthesis across multiple values, or handling missing data are not evaluated.

**No multi-turn evaluation.** All evaluations use single-turn queries. Real financial research often involves iterative refinement, follow-up questions, and cross-referencing—capabilities not tested here.

**Dataset scale.** With 500 questions, statistical power is limited for fine-grained subgroup analyses. Confidence intervals on accuracy estimates range from $\pm 2$–4pp for overall results and wider for category-specific breakdowns.

**Cross-provider comparisons.** Comparisons across providers reflect differences in complete agent systems (model + SDK + web tools), not isolated model capabilities. This is by design: the benchmark measures what practitioners experience when deploying each provider's recommended stack. Within-provider comparisons (e.g., MCP vs WebOnly, reasoning vs base mode) control for SDK and tool differences, supporting stronger causal claims.

**Provider-specific reasoning implementations.** Extended reasoning is implemented differently across providers (Extended Thinking vs Reasoning API vs ThinkingConfig). Cross-provider comparisons of reasoning benefit should be interpreted as comparing specific implementations, not reasoning capability in general.

**Financial domain only.** Findings about tool availability, reasoning benefits, and geographic gaps may not transfer to other structured data retrieval domains (e.g., medical records, legal documents, scientific databases).

# 8 Conclusion

We present FinRetrieval, a benchmark for evaluating AI agents on financial data retrieval. The dataset comprises 500 questions with ground truth answers, agent responses from 14 configurations across three frontier providers, and complete execution traces including all tool calls.

Our evaluation reveals four key findings:

1. **Tool availability dominates performance.** Claude Opus achieves 90.8% accuracy with structured data tools but only 19.8% with web search alone—a 71pp gap that exceeds other models by 3–4×.

2. **Reasoning benefits vary inversely with base capability.** OpenAI gains +9.0pp from reasoning mode while Claude Opus gains only +2.8pp, explained by OpenAI's weaker base-mode tool utilization rather than superior reasoning.

3. **First-query success drives efficiency metrics.** Correct answers use fewer tool calls because first-query success (87% for Claude) enables the optimal 3-call path with 93% accuracy; failure triggers extended search loops with 77% accuracy.

4. **Geographic gaps reflect data conventions.** The 5.6pp US advantage for Claude stems from fiscal year naming mismatches for non-December FYE companies, not inherent model limitations.

These findings suggest that improving financial AI systems should prioritize structured data integration over reasoning enhancements or model selection. Beyond tool availability, the path from 90% to higher accuracy lies in tool description completeness: 63% of remaining errors stem from undocumented conventions that explicit tool specifications could address. The open dataset enables further research on agent behavior, error patterns, and the development of more robust financial retrieval systems. Future work includes multi-turn evaluation protocols and integration with additional financial data sources.

The dataset, evaluation code, and complete tool traces are available at huggingface.co/datasets/daloopa/finretrieval, with schema documentation (Appendix B) and reproduction instructions (Appendix D).

# 9 Acknowledgments

# 10 Appendix

## 10.1 Appendix A: Example Questions

The dataset contains 500 questions spanning six financial categories. Below are representative examples illustrating the diversity of metrics, periods, and geographic coverage.

### 10.1.1 Representative Questions by Category

**Income Statement** — Stryker Corp (SYK, USA)

*Question*: What was Stryker Corporation's amortization of intangible assets within operating expenses for the full fiscal year 2020, in USD millions?

*Answer*: $472 million

**Balance Sheet** — Covestro AG (XTRA:1COV, Germany)

*Question*: What was the balance of inventories for raw materials and supplies on Covestro AG's consolidated balance sheet as of fiscal year-end Dec. 31, 2020, in EUR millions?

*Answer*: €537 million

**Cash Flow** — Metro Inc (MRU:CN, Canada)

*Question*: In fiscal 2018, what was Metro Inc.'s cash flow from investing activities related to the equity forward transaction on the investment at fair value, in CAD millions?

*Answer*: CAD 68.4 million

**Guidance** — Socionext Inc (6526.T, Japan)

*Question*: During its calendar Q1 2024 earnings announcement, what ordinary income did Socionext forecast for the consolidated fiscal year ending March 31, 2025, in JPY millions?

*Answer*: ¥27,000 million

**Segments** — Vasta Platform Ltd (VSTA, Brazil)

*Question*: What was Vasta Platform Ltd.'s net revenue from sales and services for the Content & EdTech Platform segment in fiscal year 2021, in thousands of Brazilian reais?

*Answer*: BRL 850,713 thousand

**Operational KPIs** — Imax Corp (IMAX, USA)

*Question*: In Imax Corp's fiscal year 2018, how many IMAX Digital GT Laser 3D theatre systems were included in the company's new backlog as of December 31?

*Answer*: 10 units

### 10.1.2 Period Confusion Examples

Period confusion accounts for 43% of agent errors across all configurations (63% for the best-performing Claude Opus with reasoning; see Section 6.1). The following examples illustrate the two most common patterns.

**Pattern 1: Announcement vs Target Period (Guidance)**

| Interpretation | Calendar Months | Result |
|---|---|---|
| Calendar Q4 2015 | Oct–Dec 2015 | ✘ Wrong period |
| Fiscal Q4 2015 (April FYE) | Jan–Mar 2015 | ✓ Correct |

Table 9: Fiscal Q4 2015 maps to different calendar months depending on fiscal year-end date.

*Question*: When announcing its results based on information available as of March 11, 2021, what was the low end of Domo, Inc.'s revenue guidance in USD millions for Q1 fiscal 2022?

*Expected Answer*: $56.5 million

*Common Error*: Agent queries period "2022Q1" (the forecasted quarter) and retrieves $60 million—guidance announced in a *later* earnings call. The correct answer requires querying "2020FY" (the announcement period), because guidance is stored under when it was disclosed, not the period being forecasted.

**Pattern 2: Period Convention Misalignment**

Period convention misalignment encompasses two sub-patterns: (a) quarter identity differences, where fiscal Q1 $\neq$ calendar Q1 for non-December FYE companies, and (b) year labeling conventions, where the same fiscal year may be labeled "2022FY" (start-year) or "2023FY" (end-year).

*Question*: What was Smith & Wesson Brands' loss from discontinued operations related to the security solutions division in fiscal Q4 2015, in USD thousands?

*Expected Answer*: -$53 thousand

*Common Error (Quarter Identity)*: Agent queries calendar Q4 2015 instead of recognizing that Smith & Wesson's April fiscal year-end means fiscal Q4 2015 = calendar Q1 2015 (January–March), as shown in Table 9.

**Pattern 3: Granularity Mismatch (FY vs Q4)**

*Question*: What was Marriott Vacations Worldwide Corporation's total timeshare contract sales for full fiscal year 2020, in USD millions?

*Expected Answer*: $656 million

*Common Error*: Agent queries period "2020FY" but the value is stored at Q4 granularity ("2020Q4"). Year-end totals and point-in-time balance sheet items are often stored under the final quarter rather than a separate annual period.

These three patterns correspond to the period filter failure root causes identified in Section 5.5: period convention misalignment (Pattern 2, 46%), guidance storage (Pattern 1, 23%), and granularity mismatch (Pattern 3, 27%). Period confusion disproportionately affects non-US companies (which more frequently have non-calendar fiscal years) and guidance questions (which have inherent announcement-vs-target ambiguity).

## 10.2 Appendix B: Dataset Documentation

The FinRetrieval dataset is released on HuggingFace (daloopa/finretrieval) as four Parquet files. The complete schema is documented in the dataset card; key columns are summarized below.

### 10.2.1 questions.parquet (500 rows)

| Column | Type | Description |
| --- | --- | --- |
| index | int | Question ID (0–499) |
| question | string | Natural language question |
| answer | string | Ground truth answer |
| value, unit | string | Numeric value and scale |
| category | string | Financial category |
| ticker, company | string | Company identifiers |
| fiscal_period, calendar_period | string | Period in both formats |
| period_type | string | Which period format the question uses |

### 10.2.2 responses.parquet (7,000 rows)

| Column | Type | Description |
| --- | --- | --- |
| index | int | Question ID |
| configuration | string | Agent config (e.g., opus4.5) |
| response | string? | Agent response text (null if error/timeout) |
| status | string | success, error, or timeout |
| duration_ms | int | Response time in milliseconds |
| model | string | API model identifier |
| input_tokens, output_tokens | int | Token counts |

### 10.2.3 scores.parquet (7,000 rows)

| Column | Type | Description |
| --- | --- | --- |
| index, configuration | | Join keys |
| is_correct | bool | Whether response matches ground truth |
| expected_value, expected_unit | string | Normalized ground truth |
| extracted_value, extracted_unit | string? | Values extracted from response |
| could_extract | bool | Whether extraction succeeded |
| error_reason | string? | Mismatch description if incorrect |

### 10.2.4 tool_traces.parquet (7,000 rows)

| Column | Type | Description |
| --- | --- | --- |
| index, configuration | | Join keys |
| tool_calls | string | JSON array of tool call objects |
| num_tool_calls | int | Number of tool invocations |
| total_duration_ms | int | Sum of tool execution times |

Each tool call object contains: id, name, timestamp, start_ms, duration_ms, input, output, is_error, error_message.

## 10.3 Appendix C: Scoring Prompt

The LLM judge (GPT-5.2, temperature 0) receives the following structured prompt for each response:

```
You are scoring a chatbot's response to a financial question.

# Ground Truth
Company: {company} ({ticker})
Period: {period}
Value: {value}
Unit: {unit}

Question: {question}
Answer: {answer}

---

# Task 1: Extract Chatbot's Prediction

Chatbot Response:
{chatbot_response}

Extract the numeric prediction using this MANDATORY checklist:

☐ STEP 1 - LOCATE VALUE
  - Multiple values present? → Extract value matching question's PERIOD and UNIT
  - Table AND prose present? → Prefer TABLE value (more explicit)
  - Precise AND rounded? → Match question's requested precision

☐ STEP 2 - HANDLE SIGN
  - Parentheses "(X)" → negative: "(1,234)" → "-1234"
  - Context words → negative: "outflow", "decline", "loss", "decrease"
  - Question asks about outflow + response gives magnitude with directional word → negative
  Example: "£13.5 million (cash outflow)" → extracted_value: "-13.5"

☐ STEP 3 - NORMALIZE NUMBER
  - Remove ALL commas: "158,528" → "158528"
  - Keep decimal points: "1,234.56" → "1234.56"

☐ STEP 4 - CONVERT UNITS (before comparing)
  - thousand ↔ million: multiply/divide by 1000
  - "13.359 million" = "13,359 thousand" ✓

Output:
- extracted_value: signed number without commas
- extracted_unit: scale (e.g., "Million", "Thousand", "Percent")
- currency: if present (e.g., "AED", "USD")
- could_extract: false if no numeric value could be found

---

# Task 2: Compare with Unit Awareness
```

```
CORRECT if:
- Unit-converted values match: "13.359 million" = "13,359 thousand" ✓
- Sign tolerance for directional questions with context

INCORRECT - Describe the mismatch factually:
- State observable difference (response value vs expected value)
- Include magnitude if relevant (e.g., "1000x difference", "opposite sign")
- Keep to 1 sentence, focused on observable symptoms

Output fields:
- expected.answer, expected.value, expected.unit, expected.currency
- is_correct: true if chatbot answer matches ground truth
- error.has_error: true if there is an error
- error.error_reason: symptom-focused mismatch description
```

The judge returns structured JSON output parsed via Pydantic for consistent scoring.

## 10.4 Appendix D: Reproducibility

- **Code**: github.com/daloopa/finretrieval
- **Dataset**: huggingface.co/datasets/daloopa/finretrieval

See repository README for environment setup and evaluation commands.

## 10.5 Appendix E: Sampling Methodology

### 10.5.1 Category Assignment

Series are assigned to one of six financial categories by pattern-matching the first component of their pipe-delimited hierarchy name. For example, series beginning with "Income Statement," "GAAP to Non-GAAP," or "EBITDA" map to income statement performance; series beginning with "Segmental Breakdown" or "Geographic Information" map to segments/geography. This rule-based assignment covers 97% of all series in the database; uncategorized series are excluded from sampling.

### 10.5.2 Hierarchical Series Sampling

After selecting a company and category, we sample a specific metric series from within that subset using hierarchical sampling with $\sqrt{\log(n)}$ dampening. This scoping is important: the hierarchical sampling addresses over-representation *within* categories, not across them.

**Problem.** In the source database, series names containing "breakdown" comprise 46% of all series overall, and up to 84% within specific categories (e.g., operational KPIs for mining companies). Naive uniform sampling would produce a dataset dominated by breakdown metrics.

**Algorithm.** Series names follow a pipe-delimited hierarchy (e.g., "Other breakdown | By mine | Total cash costs"). We sample hierarchically:

1. At each depth level, extract unique components
2. Weight each component by $\sqrt{\log(n+1)}$ where $n$ is the subtree size
3. Sample one component proportional to weights
4. Filter to series with chosen component and recurse
5. At leaf level, sample uniformly from remaining series

| Function | Company A | Company B | Selected |
|---|---|---|---|
| None (uniform) | 0pp | 0pp | |
| $\sqrt{n}$ | $-14.0$pp | $-10.9$pp | |
| $\log(n)$ | $-31.1$pp | $-21.0$pp | |
| $\sqrt{\log(n)}$ | **$-39.4$pp** | **$-28.6$pp** | ✓ |
| $\log(\log(n))$ | $-37.0$pp | $-26.2$pp | |

Table 10: Breakdown reduction (percentage point change from baseline) by dampening function. $\sqrt{\log(n)}$ achieves best reduction.

| Provider | API | Parameters |
|---|---|---|
| Anthropic | Extended thinking | `max_thinking_tokens=8000` |
| OpenAI | Reasoning API | `reasoning_effort="high"`, `summary="detailed"` |
| Google | ThinkingConfig | `thinking_level="high"`, `include_thoughts=True` |

Table 11: Reasoning mode configuration by provider.

**Why $\sqrt{\log(n)}$?** We evaluated multiple dampening functions:

**Result.** The $\sqrt{\log(n)}$ weighting reduces breakdown over-sampling by 28–39 percentage points while maintaining reasonable variance. Single-series branches receive 13% probability (vs 33% with uniform hierarchical sampling), preventing pathological over-weighting of rare metrics.

## 10.6 Appendix F: Experimental Details

### 10.6.1 Reasoning Mode Parameters

Extended reasoning is implemented differently across providers:

### 10.6.2 Agent SDK Versions

All agents were implemented using official SDKs:

- Claude Agent SDK (Anthropic)
- OpenAI Agents SDK
- Google ADK (Agent Development Kit)

### 10.6.3 Tool Configuration

**MCP Tools** (Model Context Protocol):

- `discover_companies`: Maps ticker/name $\to$ company ID
- `discover_company_series`: Returns available metrics filtered by keywords
- `get_company_fundamentals`: Retrieves values for company, metrics, periods

**WebSearch**:

- Claude/OpenAI: Provider-specific web search tools
- Google: Google Search via ADK

### 10.6.4 Evaluation Parameters

- Questions per configuration: 500
- Total evaluations: 7,000 (14 configurations $\times$ 500 questions)
- Timeout per question: 120 seconds

- Temperature: 0 (deterministic) for all agent responses
- Scorer: GPT-5.2, temperature 0

## 10.7 Appendix G: Finding A Decomposition

The 71pp tool availability gap for Claude decomposes into behavioral and infrastructure components.

### 10.7.1 Behavioral Component ( 35pp)

Claude's WebSearch-only configuration exhibits a 55% "gave up" rate—cases where the agent found relevant information but failed to commit to an answer. By comparison:

- Google: under 1% gave-up rate
- OpenAI: 8% gave-up rate

**Pattern**: Claude often locates the correct answer in search snippets but continues searching for "better" sources, eventually timing out or explicitly declining to answer due to uncertainty.

**Representative Case**: Question asks for Q3 2024 revenue. Claude's search returns a snippet containing "Q3 2024 revenue: $45.2 billion" but Claude responds: "I found several references to Q3 2024 revenue but cannot verify the exact figure from primary sources. I recommend checking the company's investor relations page."

### 10.7.2 Infrastructure Component ( 36pp)

WebSearch tools return only search result snippets, not full document content. This limitation affects all models but has outsized impact on Claude:

1. **PDF inaccessibility**: SEC filings (10-K, 10-Q) contain precise figures in PDF tables that WebSearch cannot access
2. **Snippet truncation**: Financial tables often exceed snippet length limits
3. **Dynamic content**: Interactive charts and data visualizations are invisible to search

**Amplification effect**: Claude's extensive search behavior (average 12.3 searches vs 4.1 for Google) amplifies infrastructure limitations—more searches means more encounters with inaccessible content, reinforcing uncertainty.

### 10.7.3 Decomposition Summary

## 10.8 Appendix H: Statistical Details

### 10.8.1 Confidence Intervals

With 500 questions per configuration, 95% confidence intervals for accuracy estimates are approximately $\pm$2–4pp depending on the base rate:

### 10.8.2 Per-Category Accuracy

Accuracy varies by financial category (Claude Opus-R shown):

| Component | Contribution | Evidence |
|---|---|---|
| Behavioral (gave-up) | 35pp | 55% gave-up rate vs <1% for Google |
| Infrastructure | 36pp | Shared across providers, amplified by search volume |
| **Total Gap** | **71pp** | |

Table 12: Decomposition of Claude's 71pp tool availability gap.

| Accuracy | Standard Error | 95% CI |
|---|---|---|
| 90% | 1.3% | ±2.6pp |
| 80% | 1.8% | ±3.5pp |
| 70% | 2.0% | ±4.0pp |
| 50% | 2.2% | ±4.4pp |
| 20% | 1.8% | ±3.5pp |

Table 13: Standard errors and 95% confidence intervals for accuracy estimates (n=500).

| Category | n | Accuracy | 95% CI |
|---|---|---|---|
| Cash Flow | 93 | 94.6% | ±4.6pp |
| Balance Sheet | 98 | 92.9% | ±5.1pp |
| Income Statement | 80 | 91.3% | ±6.2pp |
| Guidance | 73 | 89.0% | ±7.2pp |
| Segments | 64 | 87.5% | ±8.1pp |
| Operational KPIs | 92 | 86.9% | ±6.9pp |

Table 14: Accuracy by financial category for Claude Opus with reasoning.

### 10.8.3 Statistical Significance

Key comparisons and their significance (two-proportion z-test):

- **MCP vs WebOnly (Claude)**: 90.8% vs 19.8%, $z = 22.5$, $p < 0.001$
- **Reasoning benefit (OpenAI)**: 89.2% vs 80.2%, $z = 4.0$, $p < 0.001$
- **US vs Non-US (Claude)**: 93.4% vs 87.8%, $z = 2.2$, $p = 0.03$

All primary findings are statistically significant at $\alpha = 0.05$.

## 10.9 Appendix I: Error Taxonomy

Categorization of 524 incorrect responses across all eight MCP-enabled configurations, extracted from detailed per-question analysis. Scorer issues and question errors from initial analysis were resolved via manual score override or question recollection.

### 10.9.1 Agent Response Errors (96%)

Agent errors break down into subcategories:

**Period confusion** includes:
- Period convention misalignment: quarter identity differences (fiscal Q1 ≠ calendar Q1) and year labeling conventions (start-year vs end-year)—causes geographic accuracy gap (Section 5.6)
- Guidance stored under announcement period, not target period

| Category | Count | % | Description |
|---|---|---|---|
| Agent errors | 501 | 96% | Model made incorrect decision |
| Data issues | 22 | 4% | Source data mislabeling |
| System errors | 1 | <1% | API timeout/failure |

Table 15: High-level error categorization across all MCP-enabled configurations.

| Subcategory | Count | % |
| --- | --- | --- |
| Period confusion | 213 | 43% |
| Series selection | 83 | 17% |
| Field selection | 52 | 10% |
| Sign errors | 35 | 7% |
| Search strategy | 18 | 4% |
| Other | 100 | 20% |
| **Total agent errors** | **501** | **100%** |

Table 16: Agent error subcategories (percentages sum to 100% of agent errors).

- Quarterly vs annual granularity mismatch

Root cause breakdown for period filter failures is detailed in Section 5.5.

**Series selection** includes:
- Similar but incorrect metric (gross vs net, total vs component)
- Wrong hierarchy level in breakdown
- Adjacent series confusion

**Field selection** includes:
- Wrong value field (quarterly vs YTD)
- Currency misinterpretation

### 10.9.2 Data Issues (4%)

22 cases of source data problems:
- Series mislabeling in Daloopa database
- Translation errors in non-English series names
- Inverted hierarchy labels

### 10.9.3 System Errors (<1%)

Only 1 API timeout across 524 failures, indicating robust infrastructure.

Detailed root cause analysis for the best-performing configuration (Claude Opus with reasoning, 46 agent errors) is presented in Section 6.1.

## 10.10 Appendix J: Error Analysis Details

### 10.10.1 Common Error Patterns

**Period Confusion.** The most frequent error pattern involves period convention misalignment—both quarter identity differences (fiscal Q1 $\neq$ calendar Q1) and year labeling conventions (start-year vs end-year). Example: Question asks for "fiscal Q1 2024" but agent queries "Q1 2024" without fiscal qualifier, retrieving calendar Q1 data for companies with non-December fiscal year-ends.

**Series Selection Errors.** Agents sometimes retrieve similar but incorrect metrics. Example: Question asks for "net revenue" but agent retrieves "total revenue" or "revenue from operations"—related but distinct line items in financial statements.

**Sign Handling.** Financial data uses multiple conventions for negative values: parentheses "(1,234)", explicit minus "−1234", or context words ("outflow", "loss"). Agents occasionally misinterpret or drop signs during extraction.

**Over-Calculation.** Particularly in reasoning mode, agents sometimes attempt to "correct" retrieved values by performing unnecessary calculations. Example: Agent retrieves correct fiscal H1 value but then pro-rates it across calendar boundaries, introducing errors.

### 10.10.2 Model-Specific Patterns

**Claude**: Higher gave-up rate in WebSearch mode; more conservative about committing to answers without verification.

**GPT**: Lower first-DS success rate in base mode; fewer exploration calls before committing.

**Gemini**: Over-calculation pattern more prevalent in reasoning mode; tendency to second-guess retrieved values.